\pdfoutput=1
\documentclass[10pt,twocolumn]{article}

\usepackage[margin=0.72in,columnsep=0.24in]{geometry}
\usepackage{graphicx}
\usepackage{txfonts}
\usepackage{booktabs}
\usepackage{natbib}
\usepackage{xurl}
\usepackage{dblfloatfix}
\usepackage{hyperref}
\hypersetup{
  colorlinks=true,
  linkcolor=blue,
  citecolor=blue,
  urlcolor=blue,
  pdftitle={Testing the recovery of the NGC 6505 Einstein ring in deep OSN 1.5-m imaging},
  pdfauthor={Homer Davila Gutierrez et al.}
}

\newcommand{\arcsec}{\ensuremath{^{\prime\prime}}}
\newcommand{\arcmin}{\ensuremath{^{\prime}}}
\newcommand{\degr}{\ensuremath{^{\circ}}}
\newcommand{\tablefoot}[1]{\par\vspace{0.5ex}{\footnotesize\textit{Notes.} #1}}

\begin{document}

\twocolumn[
\begin{@twocolumnfalse}
\begin{center}
{\LARGE\bfseries Testing the recovery of the NGC 6505 Einstein ring in deep OSN 1.5-m imaging\par}
\vspace{1.15em}
{\normalsize
\mbox{Homer D\'avila Guti\'errez$^{1,*}$}, \mbox{Julio Carlos Bertua Marasca$^{2,4}$}, \mbox{Alejandro Heredia Teva$^{3}$},\\[0.35em]
\mbox{Irene Naranjo Arroyo$^{3}$}, \mbox{Fernando N\'u\~nez Mart\'inez$^{2}$}, \mbox{Miguel Ruiz Traves\'i$^{3}$}, \mbox{Eugenio Grande S\'aez$^{2}$}\par}
\vspace{0.9em}
{\small
$^{1}$SKYCR.ORG, San Jos\'e, Costa Rica \quad $^{*}$Corresponding author: \href{mailto:cosmos@skycr.org}{cosmos@skycr.org}\\
$^{2}$Universidad Internacional de la Rioja (UNIR), Logro\~no, Spain\\
$^{3}$Independent researcher, Spain\\
$^{4}$Independent researcher, Buenos Aires, Argentina\par}
\end{center}
\vspace{0.75em}

{\large\bfseries Abstract\par}
\vspace{0.35em}
\noindent
The discovery of a complete Einstein ring around the nearby elliptical galaxy NGC 6505 ($z = 0.042$) by the Euclid mission provides a rare, precisely characterised low-redshift strong lens. Its Einstein radius, $\theta_{\rm E} = 2.500\arcsec$, is comparable to the angular resolution attainable from typical mid-aperture ground-based facilities, which makes the recoverability of the ring from the ground an open practical question. We test whether the ring can be recovered in deep Cousins-R imaging of NGC 6505 obtained with the 1.5-m T150 telescope at the Sierra Nevada Observatory (OSN), totalling 8.03 h of integration at a final effective seeing of $\mathrm{FWHM} = 2.47\arcsec$ ($\theta_{\rm E}/\mathrm{FWHM} = 1.01$). After subtracting a non-parametric isophotal model of the galaxy, we search the residual for coherent arc- or ring-like structure and find none at the Euclid-measured Einstein radius, with a maximum value of $1.5\sigma$ for the adopted radial recovery metric. An injection--recovery experiment using a simplified, Euclid-informed synthetic ring, processed through the same isophotal-modelling and residual-analysis pipeline, yields a maximum recovered significance between $0.42\sigma$ and $1.86\sigma$ across three PSF descriptions, below our operational $3\sigma$ recovery threshold in every case. We interpret this as a dataset-specific non-recovery driven by the $\theta_{\rm E} \approx \mathrm{FWHM}$ regime of these observations, rather than as a general detectability law.

\vspace{0.55em}
\vspace{1.15em}
\end{@twocolumnfalse}
]

\section{Introduction}

Strong gravitational lensing provides a direct probe of the projected mass distribution of foreground galaxies, independent of their luminous content \citep{Schneider1992}. Low-redshift lenses are particularly valuable because their Einstein radii subtend small physical scales and probe the innermost, baryon-dominated regions of the lens galaxy \citep{Collett2018}, but such systems are intrinsically rare and only a handful have historically been known.

This landscape changed with the discovery of a complete Einstein ring around the nearby galaxy NGC 6505 ($z = 0.042$) by the Euclid mission \citep{ORiordan2025}. The system has a precisely measured Einstein radius $\theta_{\rm E} = (2.500 \pm 0.001)\arcsec$ (corresponding to $\approx 2.1$ kpc at the lens redshift) and a total lensed source magnitude $I_{\rm E} = 18.1$ in the Euclid VIS band, making it one of the highest signal-to-noise optical strong-lens observations to date. Several additional low-redshift systems are expected to be found over the Euclid Wide Survey, which motivates asking what role ground-based facilities can play in following them up.

The difficulty is one of angular scale. With $\theta_{\rm E} = 2.5\arcsec$, the ring is comparable in size to the point spread function (PSF) delivered by mid-aperture telescopes under typical seeing, and it sits on top of the bright, smooth stellar halo of the lens galaxy. NGC 6505 is therefore an ideal controlled test case: its ring properties are known independently from Euclid, so its recoverability from the ground can be assessed against a well-defined target rather than an assumed model.

In this work we test whether the NGC 6505 ring can be recovered in deep ground-based imaging obtained with the 1.5-m T150 telescope at the OSN. Our aims are (i) to build a smooth model of the galaxy light and search the residual for coherent arc- or ring-like structure at the Euclid-measured Einstein radius, and (ii) to inject a simplified, Euclid-informed synthetic ring into the same data and process it through the same isophotal-modelling and residual-analysis pipeline, quantifying whether it would be recovered above an operational $3\sigma$ recovery threshold. We restrict our conclusions to the specific observing conditions, reduction and modelling strategy of these data; we do not attempt to derive a general detectability law.

\section{Observations and image preparation}

Imaging of NGC 6505 was obtained with the 1.5-m T150 Ritchey--Chr\'etien telescope of the Sierra Nevada Observatory (OSN, Granada, Spain), equipped with an Andor Ikon-L $2048\times2048$ CCD providing a plate scale of $0.232\arcsec\,\mathrm{pix}^{-1}$ and a field of view of $\sim7.9\arcmin\times7.9\arcmin$. Observations were carried out in the Cousins-R band on two photometric nights, 16 and 17 May 2025, yielding 18 and 26 individual frames for a total accumulated integration of 28,900 s (8.03 h). The target was observed near transit with airmass between 1.05 and 1.30, and all frames were guided to ensure sub-pixel registration. The observing log is given in Table~\ref{tab:observing_log}.

\begin{table}[!t]
\caption{Log of OSN T150 observations of NGC 6505}
\label{tab:observing_log}
\centering
\scriptsize
\resizebox{\columnwidth}{!}{%
\begin{tabular}{lccccc}
\toprule
Date (UT) & Filter & $N_{\rm exp}$ & $t_{\rm int}$ (s) & $\langle t_{\rm exp}\rangle$ (s) & Mean airmass \\
\midrule
2025-05-16 & R & 18 & 12\,750 & 708 & 1.10 \\
2025-05-17 & R & 26 & 16\,150 & 621 & 1.12 \\
Combined   & R & 44 & 28\,900 & 657 & -- \\
\bottomrule
\end{tabular}}
\tablefoot{$N_{\rm exp}$ is the number of individual frames per night, $t_{\rm int}$ is the total integration time, and $\langle t_{\rm exp}\rangle$ is the mean per-frame exposure time. The last row corresponds to the weighted combination used in the analysis.}
\end{table}

Individual frames were bias- and flat-field-corrected and combined into nightly master stacks with DeepSkyStacker, which applies per-pixel kappa-sigma clipping for cosmic-ray rejection and normalises the output to the [0,1] dynamic range. The two nightly stacks were aligned to a common frame using the asterism-matching transformation of \texttt{astroalign} \citep{Beroiz2020}, retaining 98.6\% of the common footprint, and combined into a single master image with weights proportional to their integration times. The analysis was performed on this final normalised stack; the relative flux scaling required for the injection experiment (Sect.~5) is tied to the external photometric calibration described below. A $440\times440$ pixel ($102\arcsec\times102\arcsec$) cutout was extracted around the galaxy centroid, determined by 2D Gaussian centroiding to better than 0.1 pix.

Because the frame combination discards the absolute flux scale, all photometry is referenced to an externally derived zero point. We calibrated the master image against 12 Pan-STARRS DR1 r-band stars in the field \citep{Chambers2016}, obtaining $\mathrm{ZP}_{r} = (16.15 \pm 0.05)$ mag, where the uncertainty is the scatter of the individual calibrators. We note that the observations were taken in Cousins R while the reference catalogue is Pan-STARRS r; no explicit colour term is applied, and the zero point is used only to set the flux scale of the injected synthetic ring, not to derive science-grade surface-brightness limits.

\section{PSF characterisation}

\begin{figure*}[!t]
\centering
\includegraphics[width=0.96\textwidth]{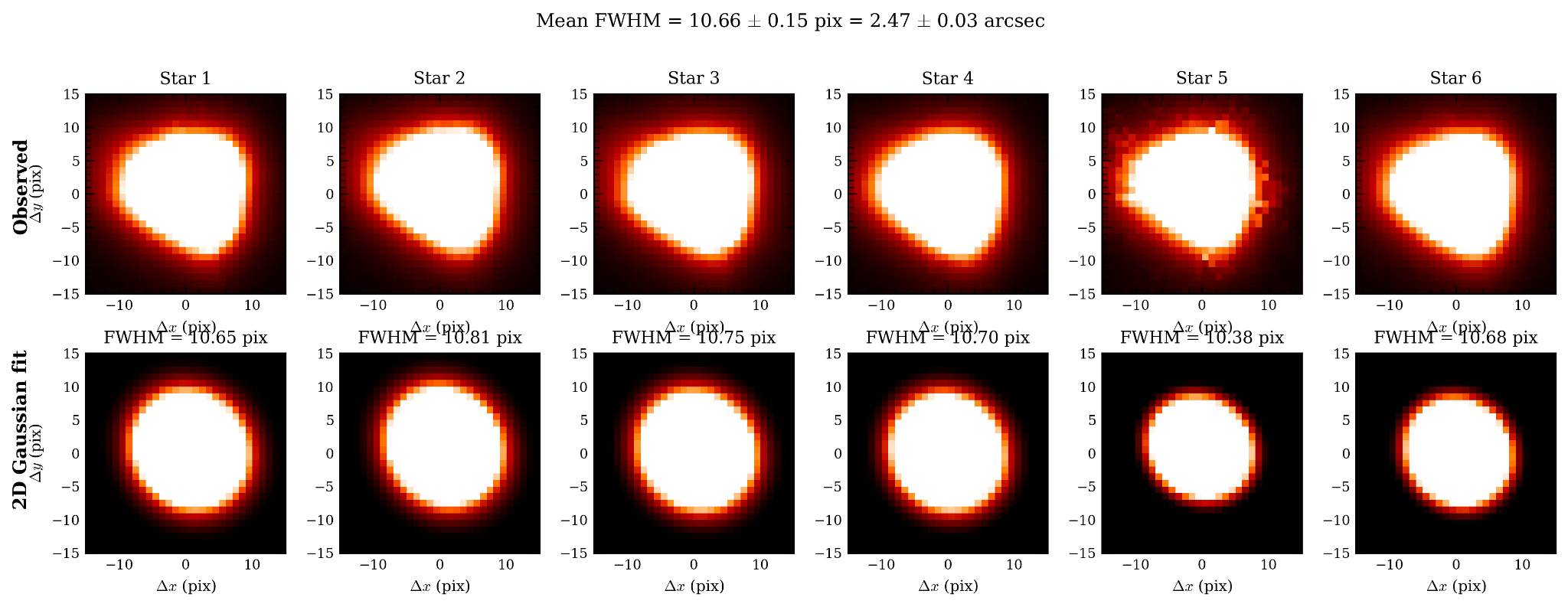}
\caption{PSF characterisation. Observed stamps of the six unsaturated field stars used to characterise the PSF of the combined image (top), and their best-fitting two-dimensional circular Gaussian models (bottom). The mean FWHM is $2.47\arcsec$; the quoted $0.03\arcsec$ uncertainty is the dispersion across the six stars.}
\label{fig:psf}
\end{figure*}

The angular resolution of the combined image was characterised by fitting the PSF to six unsaturated field stars, selected to lie in the linear regime of the detector, to be approximately circular, and to be more than 250 pixels from NGC 6505. Each star was fitted with a two-dimensional circular Gaussian, and the full width at half maximum was derived from the geometric mean of the orthogonal widths. The six measurements yield
\begin{equation}
\mathrm{FWHM} = (2.47 \pm 0.03)\arcsec\;[(10.66 \pm 0.15)\,\mathrm{pix}],
\end{equation}
where the uncertainty is the dispersion across the sample. The PSF is approximately uniform across the central region, with no significant gradient in FWHM. Inspection of the fit residuals reveals a mild triangular (trefoil) asymmetry in the stellar profiles, characteristic of low-order optical aberrations in Ritchey--Chr\'etien systems (Fig.~\ref{fig:psf}). To assess the sensitivity of our results to this feature, we later repeat the injection--recovery experiment with a Moffat profile and an empirical PSF in addition to the Gaussian baseline (Sect.~5).

The key observational quantity for this study is the ratio between the Euclid-measured Einstein radius and our PSF FWHM,
\begin{equation}
\theta_{\rm E}/\mathrm{FWHM} = 2.500\arcsec/2.47\arcsec = 1.01,
\end{equation}
which places these observations in the regime of critical PSF smearing, where the angular scale of the lensing feature is essentially equal to the resolution of the imaging.

\section{Isophotal modelling and residual search}

\subsection{Smooth galaxy model}

To search for a low-contrast residual associated with the Einstein ring we first construct a smooth model of the lens galaxy light. Contaminating field sources were identified with \texttt{photutils.detect\_sources} \citep{Bradley2024} using a $5\sigma$ threshold, and the resulting mask was dilated to cover the wings of bright contaminants; the central galaxy was explicitly excluded from the mask. A constant background, measured in an outer annulus free of galaxy light, was subtracted, and its dispersion $\sigma_{\rm bg}$ defines the noise reference used in the significance estimates below.

We fitted concentric elliptical isophotes to the masked, background-subtracted cutout using \texttt{photutils.isophote.Ellipse} \citep{Bradley2024}, which implements the iterative Fourier-mode algorithm of \citet{Jedrzejewski1987}. The fit converged over semi-major axes from 0.5 to $35\arcsec$. In the radial range surrounding the Einstein radius we measure a stable geometry, with ellipticity $\epsilon = 0.154 \pm 0.003$ and position angle $\mathrm{PA} = (134.4 \pm 2.7)\degr$; these values are used only to confirm that the subtraction model follows the local galaxy geometry, and are consistent with the range reported by \citet{ORiordan2025} from Euclid VIS imaging. We then built a non-parametric smooth model from the converged isophotes using \texttt{photutils.isophote.build\_ellipse\_model}. This approach reproduces the isophotal twist and low-order geometric variations of the galaxy, so that departures from the smooth elliptical model remain in the residual together with noise and model-subtraction systematics. Potential lensing features would therefore be searched for as spatially coherent non-elliptical residuals.

\begin{figure*}[!t]
\centering
\includegraphics[width=0.92\textwidth]{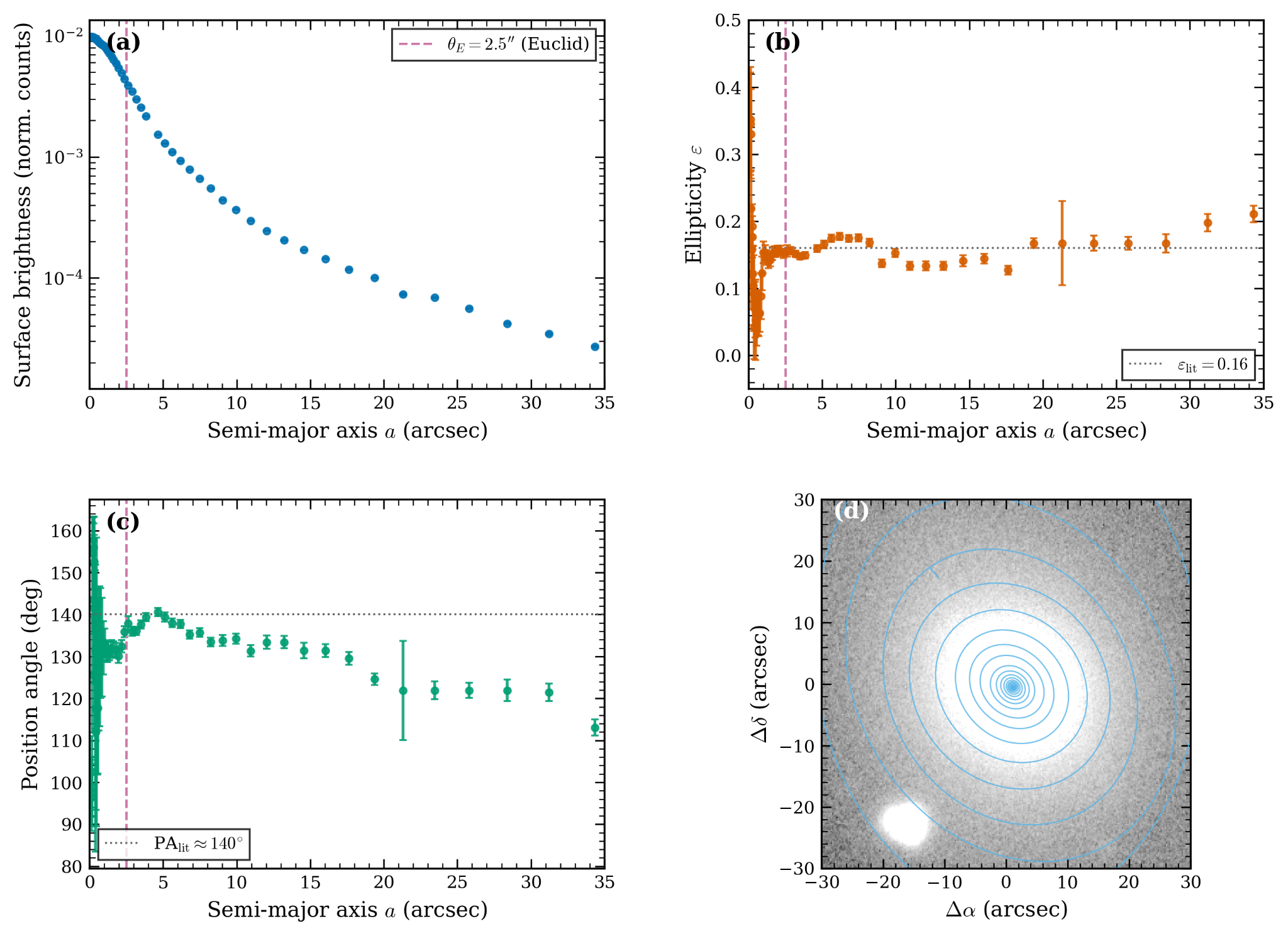}
\caption{Isophotal characterisation of NGC 6505. (a) Surface-brightness profile; (b) ellipticity profile; (c) position-angle profile; (d) background-subtracted image with the fitted isophotes overlaid. The dashed red line marks the Euclid Einstein radius $\theta_{\rm E} = 2.5\arcsec$.}
\label{fig:isophotes}
\end{figure*}

\subsection{Two-dimensional residual}

\begin{figure*}[!t]
\centering
\includegraphics[width=0.96\textwidth]{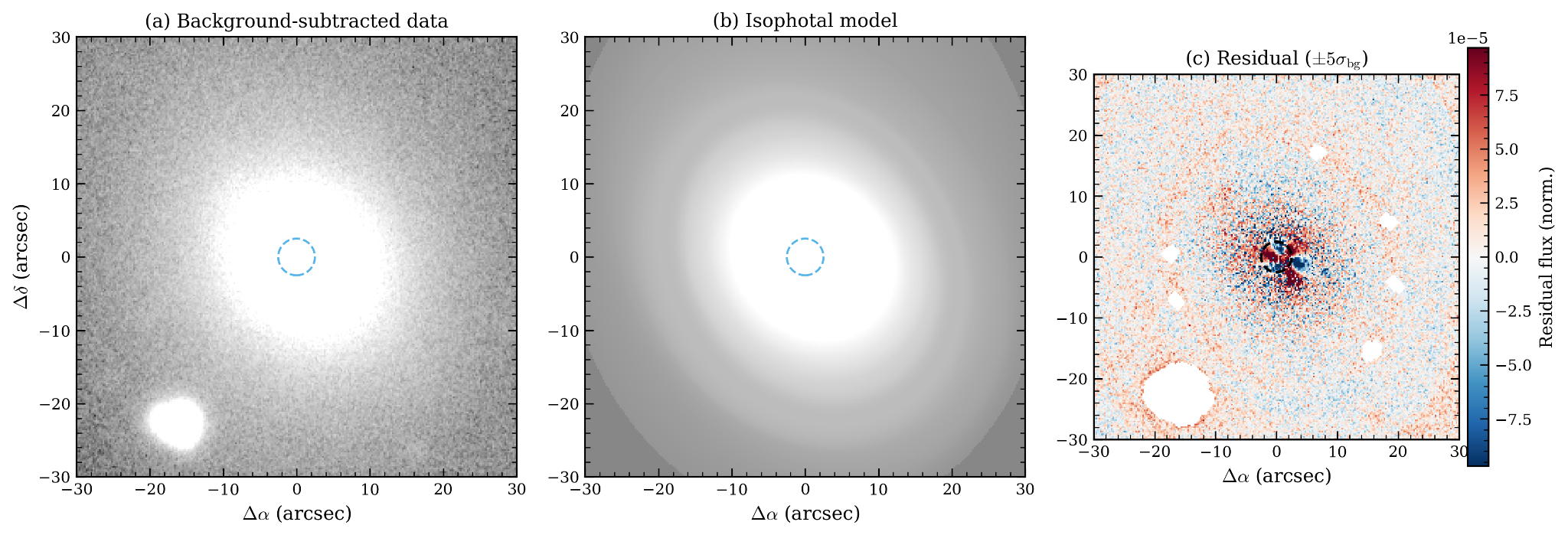}
\caption{Left: background-subtracted Cousins-R image of NGC 6505. Centre: smooth non-parametric isophotal model. Right: residual $R = D - M$, on a symmetric $\pm5\sigma_{\rm bg}$ scale. The dashed circle marks the Euclid-measured Einstein radius $\theta_{\rm E} = 2.5\arcsec$; masked contaminants appear as white regions. No coherent ring-like structure is present in the Einstein-ring region.}
\label{fig:residual}
\end{figure*}

The residual image is defined as
\begin{equation}
R(x,y) = D(x,y) - M(x,y),
\end{equation}
where $D(x,y)$ is the background-subtracted data. Figure~\ref{fig:residual} shows the data, model and residual. The full-field residual is dominated by background fluctuations and low-level model-subtraction structure, with localised features only near imperfectly subtracted bright contaminants and a low-amplitude central pattern consistent with the residual PSF asymmetry. Critically, no coherent arc-like or ring-like residual is visible at the angular scale of the Einstein ring within the displayed dynamic range.

\subsection{Radial residual profile}

To quantify the residual we azimuthally averaged it in concentric annular bins of width $\Delta r \approx \mathrm{FWHM}/4$. For each bin we computed the robust (sigma-clipped) median residual and estimated its uncertainty as $\varepsilon = 1.253\,\sigma_{\rm loc}/\sqrt{N_{\rm pix}}$, where $\sigma_{\rm loc}$ is the sigma-clipped standard deviation within the bin, $N_{\rm pix}$ the number of pixels it contains, and the factor 1.253 corrects the standard error for a median estimator on Gaussian data. We adopted a per-bin metric defined as the ratio of the median to this uncertainty. We stress that, because the image is convolved, resampled and stacked, adjacent pixels and bins are correlated; we therefore treat this quantity not as an absolute Gaussian significance but as an internal comparison metric, applied identically to the observed residual and to the injection--recovery tests of Sect.~5, and we refer to the $3\sigma$ level as an operational recovery threshold rather than a formal detection probability. This ensures that data and injections are compared on the same footing.

\begin{figure*}[!t]
\centering
\includegraphics[width=0.72\textwidth]{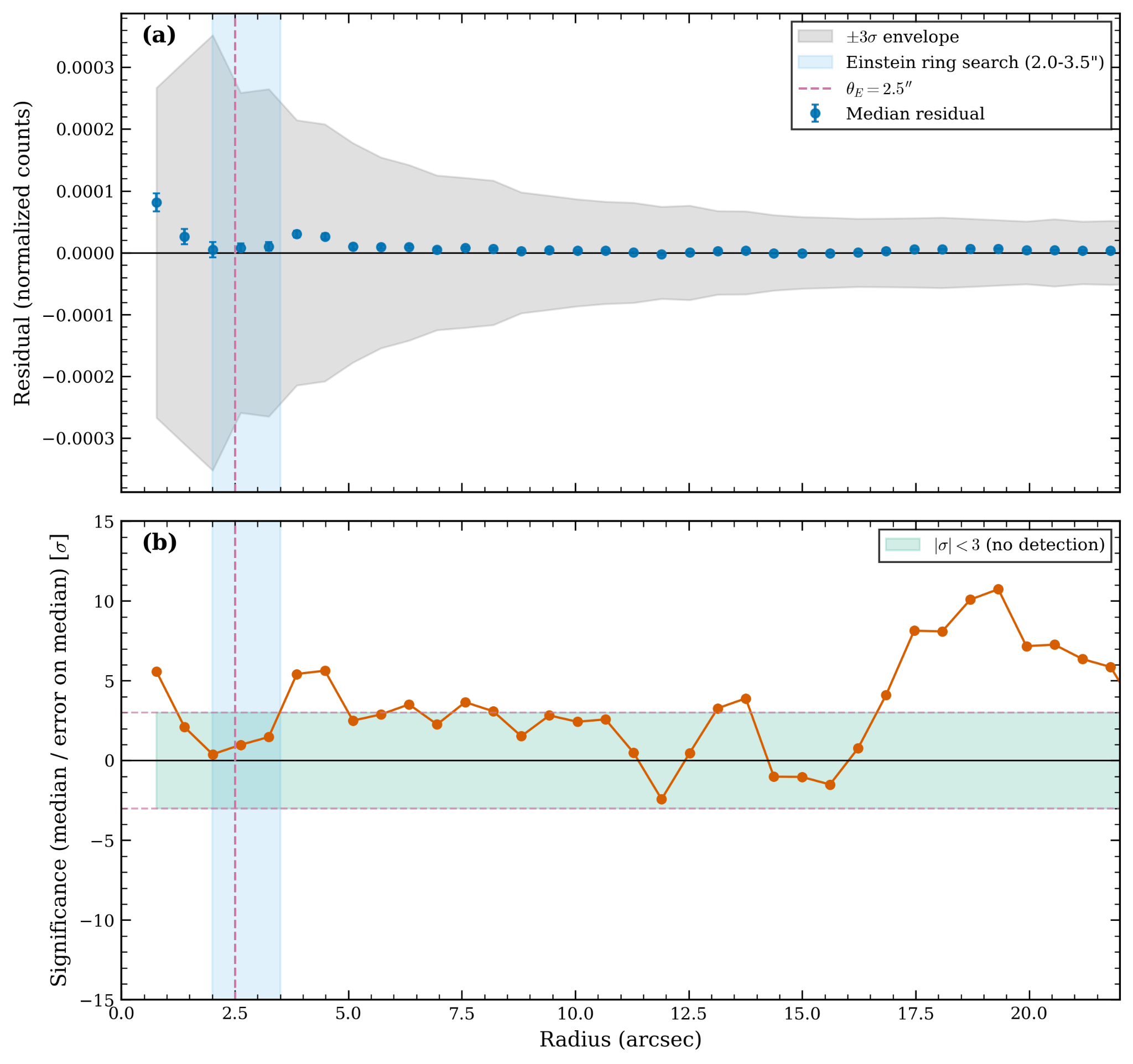}
\caption{Radial residual profile. Top: median residual in concentric annular bins as a function of radius, with the local fluctuation envelope. Bottom: per-bin significance. The vertical band marks the Einstein-ring search range ($2.0$--$3.5\arcsec$) and the dashed line the Euclid-measured $\theta_{\rm E} = 2.5\arcsec$. No bin in the search range exceeds the adopted $3\sigma$ level; the maximum significance there is $1.5\sigma$.}
\label{fig:radial}
\end{figure*}

Within the adopted Einstein-ring search range ($2.0\arcsec < r < 3.5\arcsec$ surrounding the Euclid-measured $\theta_{\rm E}$; Fig.~\ref{fig:radial}), the median residual is consistent with zero, with a maximum bin significance of $1.5\sigma$. We therefore find no coherent residual signal at the angular scale of the Einstein ring. Positive excursions at larger radii ($r \gtrsim 17\arcsec$) are unrelated to the lensing system and coincide with the outer isophotal twist visible in our own profile (Fig.~\ref{fig:isophotes}) -- a common feature of evolved early-type galaxies \citep{Hao2006} -- and with the extended halos of bright masked contaminants.

\section{Injection--recovery experiment}

The residual search establishes that no ring is seen in these data, but does not by itself quantify whether the known NGC 6505 ring would be recovered by our pipeline. To address this, we injected a simplified synthetic ring into the final reduced cutout, convolved with our measured PSF, and processed it through the same isophotal-modelling and residual-analysis pipeline used for the data.

\subsection{Simplified Euclid-informed synthetic model}

We constructed a simplified synthetic representation informed by the published Euclid geometry and total brightness of the system \citep{ORiordan2025}, rather than a faithful reconstruction of the resolved ring. The model comprises four bright, unresolved lensed images at the reported angular positions, connected by a thin low-surface-brightness arc at $\theta_{\rm E} = 2.500\arcsec$, with a total flux set to match the Euclid measurement after an approximate colour scaling to our observed band. The complete component parameters (relative magnifications, intrinsic widths and the flux partition between images and arc) are provided with the analysis scripts in the associated Zenodo repository; we verified that the qualitative outcome is insensitive to these choices at the level explored.

\subsection{Injection and recovery procedure}

\begin{figure*}[!t]
\centering
\includegraphics[width=0.96\textwidth]{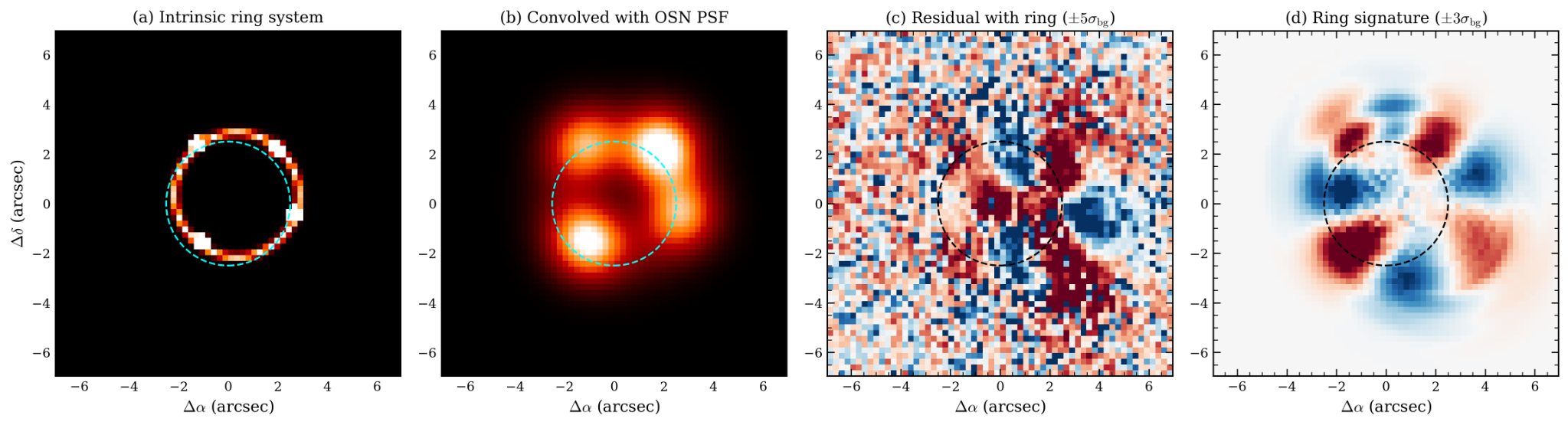}
\caption{Injection--recovery experiment with a simplified, Euclid-informed synthetic ring. From left to right: intrinsic synthetic ring (four unresolved images plus connecting arc at $\theta_{\rm E} = 2.5\arcsec$, total flux matched to the Euclid measurement); the same ring convolved with our OSN T150 PSF; residual after re-running the isophotal-modelling and residual-analysis pipeline with the injected ring; and the differential map isolating the ring imprint. The injected ring leaves a faint four-lobed signature, whose maximum radial significance does not exceed $1.86\sigma$ for any of the PSF models tested. The dashed circle marks $\theta_{\rm E} = 2.5\arcsec$.}
\label{fig:forward}
\end{figure*}

The synthetic ring was convolved with our measured PSF and added to the background-subtracted cutout. The composite image was then passed through the same isophotal-fitting pipeline as the real data, including isophote refitting and construction of a new smooth model, and the difference of the resulting residual with respect to the ring-free residual isolates the imprint of the injected ring (Fig.~\ref{fig:forward}). The convolution illustrates the dramatic loss of spatial coherence in the $\theta_{\rm E} \approx \mathrm{FWHM}$ regime: the four discrete images blend into an extended elliptical excess, most of whose light is redistributed toward the galaxy centre and reabsorbed by the isophotal model.

\subsection{Recovery result and PSF sensitivity}

We quantified the recovery using the same radial significance metric as in Sect.~4, applied to the difference between the injected and ring-free residuals. To verify that the outcome does not depend on the assumed PSF, we repeated the experiment for three descriptions: the Gaussian baseline, a Moffat profile, and an empirical PSF built directly from the field stars (Table~\ref{tab:psf_models}). Across the three models, the maximum recovered significance in the Einstein-ring search range spans $0.42\sigma$ to $1.86\sigma$, remaining below the operational $3\sigma$ recovery threshold in every case. Notably, the empirical PSF -- which retains the trefoil asymmetry of the real stellar profiles -- gives the lowest value, so that a more realistic PSF makes the ring less recoverable, not more. The non-recovery is therefore not an artefact of the Gaussian approximation.

\begin{table}[t]
\caption{Forward-modelling significance for three PSF models}
\label{tab:psf_models}
\centering
\resizebox{\columnwidth}{!}{%
\begin{tabular}{lccc}
\toprule
PSF model & FWHM (arcsec) & Max significance & Mean significance \\
\midrule
Gaussian (baseline) & 2.47 & $1.58\sigma$ & $+0.15\sigma$ \\
Moffat ($\beta = 5.3$) & 2.32 & $1.86\sigma$ & $+0.95\sigma$ \\
Empirical (EPSF) & -- & $0.42\sigma$ & $-0.01\sigma$ \\
\bottomrule
\end{tabular}}
\tablefoot{The injected ring corresponds to the Euclid-measured brightness ($I_{\rm E} = 18.1$). All three PSF models yield a maximum significance well below the $3\sigma$ detection threshold.}
\end{table}

\section{Discussion and limitations}

The combination of the direct residual search and the injection--recovery experiment yields a single, consistent result: for these data neither the observed residual nor a Euclid-informed synthetic ring processed through the same pipeline produces a $3\sigma$ recovery at the Einstein radius. The physical origin is the $\theta_{\rm E} \approx \mathrm{FWHM}$ regime, in which two effects act together. First, PSF convolution redistributes most of the ring light toward $r < \theta_{\rm E}$, where it is absorbed by the smooth galaxy model and never enters the search annulus. Second, within the annulus the residual is structured as a positive--negative pattern (Fig.~\ref{fig:forward}) whose azimuthal average partially cancels. Both effects depend explicitly on $\theta_{\rm E}/\mathrm{FWHM}$.

Several systematics were considered and none is large enough to change this conclusion for the present dataset. The photometric zero point ($\sim0.05$ mag) and the approximate colour scaling of the injected source ($\sim0.2$ mag) are sub-dominant; varying the source brightness and the image/arc flux partition over plausible ranges keeps the maximum recovered significance below the operational threshold. The explored ranges were a colour offset of $I_{\rm E} + 0.0$ to $I_{\rm E} + 0.5$ mag and an image/arc flux partition of 50--90\%, with the corresponding tests available in the Zenodo repository. The choice of PSF model, examined explicitly above, likewise does not alter the outcome. Because the image is convolved and stacked, we have deliberately used the radial significance only as an internal comparison metric between data and injections, rather than as an absolute detection statistic or as the basis for formal surface-brightness limits.

We emphasise the scope of this result. The present analysis does not establish a universal resolution threshold for ground-based Einstein-ring detection. It demonstrates that, for the depth, PSF, image reduction, galaxy-subtraction model and recovery statistic used here, neither the observed data nor the simplified Euclid-informed injection yields a $3\sigma$ recovery. The result should therefore be read as a dataset-specific recovery test rather than a general detectability law. We also note that indirect avenues -- spectroscopic identification of the lensed source, or dynamical inference -- are largely independent of imaging resolution and remain viable regardless of $\theta_{\rm E}/\mathrm{FWHM}$; our test concerns direct imaging recovery only.

\section{Conclusions}

We have carried out a controlled ground-based recovery test of the NGC 6505 Einstein ring using 8.03 h of deep Cousins-R imaging with the 1.5-m T150 telescope at the Sierra Nevada Observatory, at a final effective seeing of $\mathrm{FWHM} = 2.47\arcsec$ ($\theta_{\rm E}/\mathrm{FWHM} = 1.01$).

After subtracting a non-parametric isophotal model of the galaxy, we find no coherent arc- or ring-like residual at the Euclid-measured Einstein radius, with a maximum value of $1.5\sigma$ for the adopted radial recovery metric in the search range $2.0\arcsec < r < 3.5\arcsec$.

An injection--recovery experiment with a simplified, Euclid-informed synthetic ring, processed through the same isophotal-modelling and residual-analysis pipeline, yields a maximum recovered significance between $0.42\sigma$ and $1.86\sigma$ across Gaussian, Moffat and empirical PSF models, below our operational $3\sigma$ recovery threshold in every case. We attribute the non-recovery to the $\theta_{\rm E} \approx \mathrm{FWHM}$ regime of these observations. Higher-resolution imaging, or an alternative observing and modelling strategy, would be required to test whether the ring can be recovered from the ground; the present study does not attempt to derive a universal detectability threshold.

\section*{Acknowledgements}
The authors thank Dr. Roberto Baena Gall\'e for his supervision during the observations at the OSN, and staff astronomer V\'ictor Casanova for his assistance during the operation of the observatory. We thank Josu\'e Mosquera Hadatty and Aitana Delgado Cabanillas for their collaboration during the observing nights, and gratefully acknowledge the Master's Programme in Astrophysics of the Universidad Internacional de La Rioja (UNIR). This work is based on observations collected at the OSN, operated by the Instituto de Astrof\'isica de Andaluc\'ia (IAA-CSIC). We acknowledge the use of Euclid mission data products as published in \citet{ORiordan2025}, and of the Pan-STARRS1 archive and the SIMBAD and VizieR services at CDS, Strasbourg. This research made use of Astropy \citep{Astropy2022}, \texttt{photutils} \citep{Bradley2024}, \texttt{astroalign} \citep{Beroiz2020} and the standard scientific Python stack.

\section*{Data availability}
The reduced master image, the analysis notebooks, and the injection--recovery scripts underlying this work are available at Zenodo DOI \href{https://doi.org/10.5281/zenodo.21328613}{10.5281/zenodo.21328613}, and the reference Einstein-ring parameters are taken from the public data of \citet{ORiordan2025}.


\begin{thebibliography}{99}

\bibitem[Astropy Collaboration(2022)]{Astropy2022}
Astropy Collaboration 2022, ApJ, 935, 167, \href{https://doi.org/10.3847/1538-4357/ac7c74}{10.3847/1538-4357/ac7c74}

\bibitem[Beroiz et al.(2020)]{Beroiz2020}
Beroiz, M., Cabral, J. B., \& Sanchez, B. 2020, Astronomy and Computing, 32, 100384, \href{https://doi.org/10.1016/j.ascom.2020.100384}{https://doi.org/10.1016/j.ascom.2020.100384}

\bibitem[Bradley et al.(2024)]{Bradley2024}
Bradley, L., et al. 2024, \href{https://doi.org/10.5281/zenodo.596036}{10.5281/zenodo.596036}

\bibitem[Chambers et al.(2016)]{Chambers2016}
Chambers, K. C., et al. 2016, \href{https://arxiv.org/abs/1612.05560}{https://arxiv.org/abs/1612.05560}

\bibitem[Collett et al.(2018)]{Collett2018}
Collett, T. E., et al. 2018, Science, 360, 1342, \href{https://doi.org/10.1126/science.aao2469}{https://doi.org/10.1126/science.aao2469}

\bibitem[Hao et al.(2006)]{Hao2006}
Hao, C. N., et al. 2006, MNRAS, 370, 1339, \href{https://doi.org/10.1111/j.1365-2966.2006.10545.x}{https://doi.org/10.1111/j.1365-2966.2006.10545.x}

\bibitem[Jedrzejewski(1987)]{Jedrzejewski1987}
Jedrzejewski, R. I. 1987, MNRAS, 226, 747, \href{https://doi.org/10.1093/mnras/226.4.747}{https://doi.org/10.1093/mnras/226.4.747}

\bibitem[O'Riordan et al.(2025)]{ORiordan2025}
O'Riordan, C. M., et al. 2025, A\&A, 694, A145, \href{https://doi.org/10.1051/0004-6361/202453014}{https://doi.org/10.1051/0004-6361/202453014}

\bibitem[Schneider et al.(1992)]{Schneider1992}
Schneider, P., Ehlers, J., \& Falco, E. E. 1992, \textit{Gravitational Lenses} (Springer), \href{https://doi.org/10.1007/978-3-662-03758-4}{https://doi.org/10.1007/978-3-662-03758-4}

\end{thebibliography}
\end{document}